# Stark tuning of single-photon emitters in hexagonal boron nitride

*Gichang Noh,[†] Daebok Choi,[†] Jin-Hun Kim,[‡] Dong-Gil Im,[‡] Yoon-Ho Kim,[‡] Hosung Seo,[†] and*

*Jieun Lee[†*]*

[†]Department of Physics and Department of Energy Systems Research, Ajou University, Suwon, 16499, Korea, [‡] Department of Physics, Pohang University of Science and Technology (POSTECH), Pohang, 37673, Korea

* To whom correspondence should be addressed, jelee@ajou.ac.kr.

**Single-photon emitters play an essential role in quantum technologies including quantum computing and quantum communications. Atomic defects in hexagonal boron nitride ($h$-BN) have recently emerged as new room-temperature single-photon emitters in solid-state systems, but the development of scalable and tunable $h$-BN single-photon emitters requires external methods that can control the emission energy of individual defects. Here, by fabricating van der Waals heterostructures of $h$-BN and graphene, we demonstrate electrical control of single-photon emission from atomic defects in $h$-BN via the Stark effect. By applying an out-of-plane electric field through graphene gates, we observed Stark**

**shifts as large as 5.4 nm per GV/m. The Stark shift generated upon a vertical electric field suggests the existence of out-of-plane dipole moments associated with atomic defect emitters, which is supported by first-principles theoretical calculations. Further, we found field-induced discrete modification and stabilization of emission intensity, reversibly controllable with an external electric field.**

Two-dimensional (2D) materials have emerged as new candidates of building blocks for quantum devices because of their unique electronic and optical properties [1,2]. Recent reports on single-photon emitters (SPE) in transition metal dichalcogenides (TMDCs) [3-6] and hexagonal boron nitride (*h*-BN) [7-13] have drawn significant interests on 2D materials for quantum information processing. However, single photons generated from different emitters show inhomogeneous energy distribution, which would be detrimental for potential applications in quantum communications. There has been great interest in overcoming the inhomogeneity of SPE in these materials, for example, by optimizing the fabrication process [11,12] or by tuning the SPE's optical energy through applying strain [13] or using the Stark effect [14]. In the Stark effect, the optical energy of each SPE could shift by the application of an external electric field, which is advantageous for the local control of individual emitters. The method has been demonstrated for SPE in TMDC monolayer at low temperatures [14]. For practical and more versatile applications of 2D-based quantum photonics, however, it is required to integrate SPE operating at room temperature into devices with tunable electric fields.

*h*-BN is a wide bandgap semiconductor ($E_{gap}$ = 6 eV) [15], in which deep level atomic defects produce single-photon emissions at room temperature. The structure of *h*-BN consists of single

*h*-BN layers that are weakly coupled by van der Waals force in the vertical direction, appropriate for fabrication of 2D heterostructures [1]. According to previous reports, native point defects in this system are normally expected to lie within the plane [7], which could create an in-plane dipole that would be difficult to control with out-of-plane electric fields. However, for tunable and scalable *h*-BN SPE with a large density, it is highly desirable to be able to control individual atomic defects using an out-of-plane electric field. In this work, we fabricate *h*-BN/graphene heterostructures to apply an out-of-plane electric field to atomic defects in *h*-BN and observe linear Stark shifts up to 5.4 nm per GV/m. First-principles theoretical calculations provide a possible origin of the observed Stark shifts induced by out-of-plane electric fields. In addition, by applying a high gate voltage, we observe the discrete change of a defect's emitter intensity, reversibly controlled by an applied electric field. These results show the potential of *h*-BN/graphene heterostructures as tunable room-temperature quantum light sources integrated in electrical circuit assemblies for scalable solid-state quantum technologies [16].

To produce *h*-BN defect centers, we annealed physically exfoliated *h*-BN flakes with thickness ranging between 100 and 200 nm in a chamber filled with argon (P = 500 mTorr) at 800 °C for 30 minutes [7]. We then employed a micro-photoluminescence (µ-PL) measurement on annealed *h*-BN using a 532 nm continuous-wave laser to find flakes with bright and stable emitters. With *h*-BN flakes containing emitters, two graphene/*h*-BN/graphene heterostructures (device A and B: images in Supporting Information Fig. S1) are fabricated to apply a vertical electric-field as shown in Fig. 1a. In the fabrication process, polypropylene carbonate (PPC) is used to dry-transfer each few-layer (FL) graphene and *h*-BN onto 100 nm thick $SiO_2$ substrates with prepatterned Ti/Au electrodes [18]. We have chosen graphene with similar thicknesses for symmetric gating electrodes. By applying a voltage ($V_g$) between the Au electrodes contacting top and bottom graphene, a vertical electric-field (*E*) is applied to the emitters. The actual field applied is $E = \frac{(\varepsilon_{BN}+2)V_g}{3t}$, according to the Lorentz local field approximation where $\varepsilon_{BN}$ and $t$ are the permittivity and thickness of *h*-BN, respectively.

We then performed a scanning μ-PL to map out the positions of bright emitters on device A as shown in Fig. 1b. The emission peak in Fig. 1c shows the characteristic zero-phonon line (ZPL) of a defect. The second-order correlation measurement of the ZPL in the inset shows $g^{(2)}(0) = 0.3$, demonstrating the single-photon nature of the emitted light [17]. We also checked the emission intensity dependence on the excitation laser polarization and collection polarization, respectively (Fig. 1d). The linear polarization dependence is obtained, indicating the anisotropy of the defect's crystal structure [7]. Together with the emission intensity saturation with increasing laser power (Supporting Information Fig. S2), we confirm the quantum light emission characteristics of the defect center. All optical measurements are conducted at *T* = 10 K where the emission linewidth is narrow except some measurements at room temperature for testing the feasibility of our scheme in ambient conditions.

Next, we applied vertical electric fields to the SPE shown in Fig. 1 and recorded the spectrum (Fig. 2a). Surprisingly, the center wavelength shifts while applying out-of-plane electric fields, clearly showing the Stark-induced tuning of the SPE and the presence of an out-of-plane electric dipole associated with the SPE. By fitting the spectra to Lorentzian functions (Fig. 1c), we extract the center wavelength and full-width at half maximum (FWHM) of the emission as a function of electric field as shown in Fig. 2b and 2c (details in Supporting Information Fig. S3). The observed shift shows a linear dependence with a slope of 0.88 nm per GV/m. To double-check that the wavelength shift of the SPE is indeed induced by the applied electric field, we monitored the spectrum while reversing the bias direction by switching the gate and ground electrodes (Supporting Information Fig. S4). There, the observed Stark shift changes the sign, ruling out the possibility that the wavelength shift is induced by the defect on the *h*-BN surface coupled to contacting graphene. Moreover, the linewidth of the emission does not change much and the extracted $g^{(2)}(0)$ are below 0.5 in the given electric field range, demonstrating that the single photon nature of the emitter is simultaneously maintained under the application of an electric field ($g^{(2)}(\Delta t)$ in Supporting Information Fig. S5). In addition, the room temperature measurement of the same defect shows a similar Stark tuning to low temperature results (Fig. 2b dotted line), demonstrating the robustness of the SPE between 10 K and room temperature. Although linewidth broadening is observed at room temperature, the measured $g^{(2)}(0) = 0.39$ at $E = 0$ ($g^{(2)}(\Delta t)$ in Supporting Information Fig. S6).

In Fig. 3, we show various types of Stark shifts observed from different SPEs, together with corresponding $g^{(2)}(\Delta t)$ and polarization dependence measurements. Fig. 3a shows a SPE showing the same sign of the Stark shift to that shown in Fig. 2, but with a generally larger slope of 5.4 nm per GV/m. For another defect in the same *h*-BN (Fig. 3b), we observed the opposite sign of linear Stark shift compared to Fig. 3a under the same configuration of gating electrodes. The existence of a defect with such opposite sign of the linear Stark shift provides an additional evidence that the measured SPEs are not on the surface, but in the bulk of *h*-BN. Although the occurrence is lower than the linear cases, we have also observed defects with V-shaped and quadratic shaped Stark shifts as exemplified in Fig. 3c and 3d, respectively, possibly indicating the field-induced flipping of dipole orientation of defects. Detailed spectra data of each SPE can be found in Supporting Information Fig. S7. By fitting the observed Stark shifts to the following equation, we extract the dipole moment ($\mu$) and polarizability ($\alpha$) of defects in *h*-BN.

$$\Delta(\hbar\omega) = -\Delta\vec{\mu} \cdot \vec{E} - \vec{E} \cdot (\Delta\vec{\alpha}/2) \cdot \vec{E}$$

Here $\hbar$ and $\omega$ are Plank constant and photon frequency, respectively. From the fitting, we find $\Delta\mu$ of the $h$-BN SPE to be distributed in the range from –0.9 D to 0.9 D (1 D = 3.3 × $10^{-30}$ C·m). For the emitters exhibiting quadratic Stark shifts, the extracted $|\Delta\alpha|$ was within ~150 Å$^3$. This $|\Delta\alpha|$ of SPE in $h$-BN is about two orders of magnitude smaller than that observed in diamond NV centers [19]. Since polarizability is proportional to the defect center's volume, such small value of $|\Delta\alpha|$ may indicate the small volume and strongly bound electron wavefunctions in these defects.

To shed light on the origin of the linear Stark shifts observed in this system, we carried out first-principles density functional theory (DFT) calculations of potential SPEs in $h$-BN. In particular, we consider $V_NX_B$-type defects (X= C, N, O) (see Fig. 4a), which are considered as strong candidates of SPEs in $h$-BN [7, 20-25]. To examine their Stark shifts, we first adopt the widely-used $C_{2v}$ structural models of $V_NX_B$-type defects [7, 20-25]. Fig. 4b shows the calculated ZPL shift of neutral $V_NN_B$ as a function of out-of-plane electric field. We find, however, that the ZPL does not show any Stark shift. In the $C_{2v}$ model of $V_NN_B$, the N impurity resides in the $h$-BN plane. Thus, the model does not possess any permanent dipole in the out-of-plane direction due to the inversion symmetry with respect to the $h$-BN plane, leading to the absence of Stark shift upon the out-of-plane electric field.

Surprisingly, however, we found that the ground-state structure of neutral $V_NX_B$ is not the widely-used $C_{2v}$ model [7, 20-25], but a reduced-symmetry $C_{1h}$ structure. Fig. 4c compares the energy of the $C_{2v}$ and $C_{1h}$ structures of the neutral $V_NX_B$ defects (X=N,C,O), which exhibits a significant energy lowering as the impurity atom ($X_B$) moves out of the plane, breaking the inversion symmetry. We also checked that this is true for neutral $V_NX_B$ in the multi-layer *h*-BN bulk. With the inversion symmetry broken, the defects develop a permanent dipole in the out-of-plane direction and exhibit linear Stark-induced ZPL shifts as shown in Fig. 4b. We note that the new $C_{1h}$ models of the neutral $V_NX_B$ defects are able to explain the essential features of the experimentally observed linear Stark shifts of the SPEs in *h*-BN although the exact correspondence between the $V_NX_B$ defects and the SPEs in *h*-BN has not been established yet [7, 20-25]. First, we theoretically estimated |*Δ*μ| of the neutral $V_NX_B$ defects to be 0.36 D ($V_NC_B$), 0.27 D ($V_NN_B$), and 0.16 D ($V_NO_B$), which are on the same order of magnitude to our measurements. Second, there are two symmetrically equivalent structures for a $C_{1h}$ $V_NX_B$ defect: $X_B$ atom being up or down (Δ = +1 or -1 in Fig. 4c), leading to two possible out-of-plane dipole orientations, which are consistent with our observations (Fig. 3a and 3b).

For the V-shaped Stark shift in Fig. 3c, our results suggest that it may be related to structural transition between two possible defect orientations of the SPE. For the $V_NX_B$ defects, their double-well potential shown in Fig. 4c could become largely asymmetric in the presence of a strong electric field owing to the Stark coupling $(-\vec{d} \cdot \vec{E})$. This implies that, for a SPE with a shallow energy barrier, its defect structure could make a transition from one to the other upon reversing the electric field direction, leading to a flipping of its electric dipole. For the quadratic Stark shift observed in Fig. 3d, on the other hand, the dipole flipping might be induced gradually with the application of an electric field. Such quadratic dependence could arise for an emitter with nondegenerate excited states, which could be induced by strain or relative orientation of the defect to the electric field direction [19]. In addition, we found that the defects exhibiting nonlinear Stark shifts (Fig. 3c and 3d) show rather larger misalignment of the excitation and emission dipole orientation compared to the ones showing linear Stark shifts (Fig. 3a and 3b), possibly suggesting the association of multiple excitation pathways for the dipole transitions in these defects [26], which is a subject of further investigation.

Finally, in Fig. 5, we present the electrical control of a defect in a device fabricated using single top graphene and bottom Si gates (device C). The detailed device image, spatial 2D map, polarization dependence and laser power saturation of this defect is shown in Supporting Information Fig. S10. By increasing the gate voltage from 0 to 60 V, we observe gate-induced discrete modification of the emission intensity at 46 V. (Fig. 5a). A real-time monitoring of the emitter with gate shows the reversibly controlled and stabilized emission intensity across the threshold voltage (Fig. 5c). The observation can be attributed to the gate-induced charging effect, which is also observed in other solid-state emitters [3, 27]. The gradual reduction of the threshold voltage with increasing laser power further supports the gate-induced charging, aided by photodoping through laser illumination (Fig. 5d).

In summary, we have demonstrated the Stark-effect-induced energy control of atomic defects in *h*-BN by forming vertical heterostructures with graphene electrodes. The ability to induce Stark shifts up to 5.4 nm per GV/m is expected to benefit the preparation of two energetically identical single defects. From the linear Stark shifts frequently found from defects in *h*-BN, we revealed possible out-of-plane tilts in the atomic bonding angle of SPE through comparison with first-principles calculations, providing insights on the mirror symmetry breaking in SPE necessary to understand their optical properties. In addition to the field-induced energy shift, the application of electric fields is found to be potentially useful in modifying the charge state of defects. Altogether, our results on electrical control of single defects in *h*-BN show the potential of 2D materials for tunable room temperature solid-state emitters, well suited for free-space quantum communication and photonic quantum information processing.

## Methods

**Measurements**

For optical measurements, we used continuous-wave 532 nm laser for excitation source which is focused onto the sample by an objective lens (NA = 0.6) with laser power of 0.159 - 1.6 mW. The emitted light is collected by the same objective lens, passed through a longpass filter, and detected by a spectrometer equipped with CCD detector. For $g^{(2)}(\Delta t)$ measurement, we directed the emitted light through an additional bandpass filter and two avalanche photodiodes (Excelitas Technologies) connected to a time-correlated single-photon counting module (qutools, quTAU). For polarization dependent measurements, two Glan-Taylor linear polarizers and an achromatic half-wave plate are inserted in the optical path. For the electrical control, a gate voltage no more than $\pm 80$ V is applied to the defects during the experiments to prevent breakdown of the fabricated devices.

**Simulations**

We performed density functional theory (DFT) calculations with a semi-local PBE functional [28] using plane-wave basis sets (with a cutoff energy of 85 Ry), optimized norm-conserving Vanderbilt (ONCV) pseudopotentials [29,30], and the QUANTUM ESPRESSO code [31]. The primary defects that we considered are neutral $V_N X_B$ (X = C, N, B) in isolated forms. To mimic the presence of an isolated defect in a single *h*-BN layer within the periodic boundary condition, we employed a 60-atom supercell (in-plane dimension = 12.6 Å × 13.1 Å), in which a *h*-BN sheet is separated from its periodic images by a 16 Å-thick vacuum space. We sampled the Brillouin Zone with the *Γ* point only. We applied an out-of-plane electric field by using the saw-tooth potential method along with the dipole correction [32]. All the defects considered were fully relaxed. The ZPL of the $V_NX_B$ defects was calculated using the $\Delta$SCF method [33]. The defect level diagrams and the defect orbitals of the $V_NX_B$ defects are described in Supporting Information Figure S8. Numerical convergence in terms of the supercell size, the plane-wave cutoff energy, and the vacuum thickness were thoroughly examined (Supporting Information Fig. S9), and we found that the ZPL shift converges within 0.1 nm using our numerical set-ups.

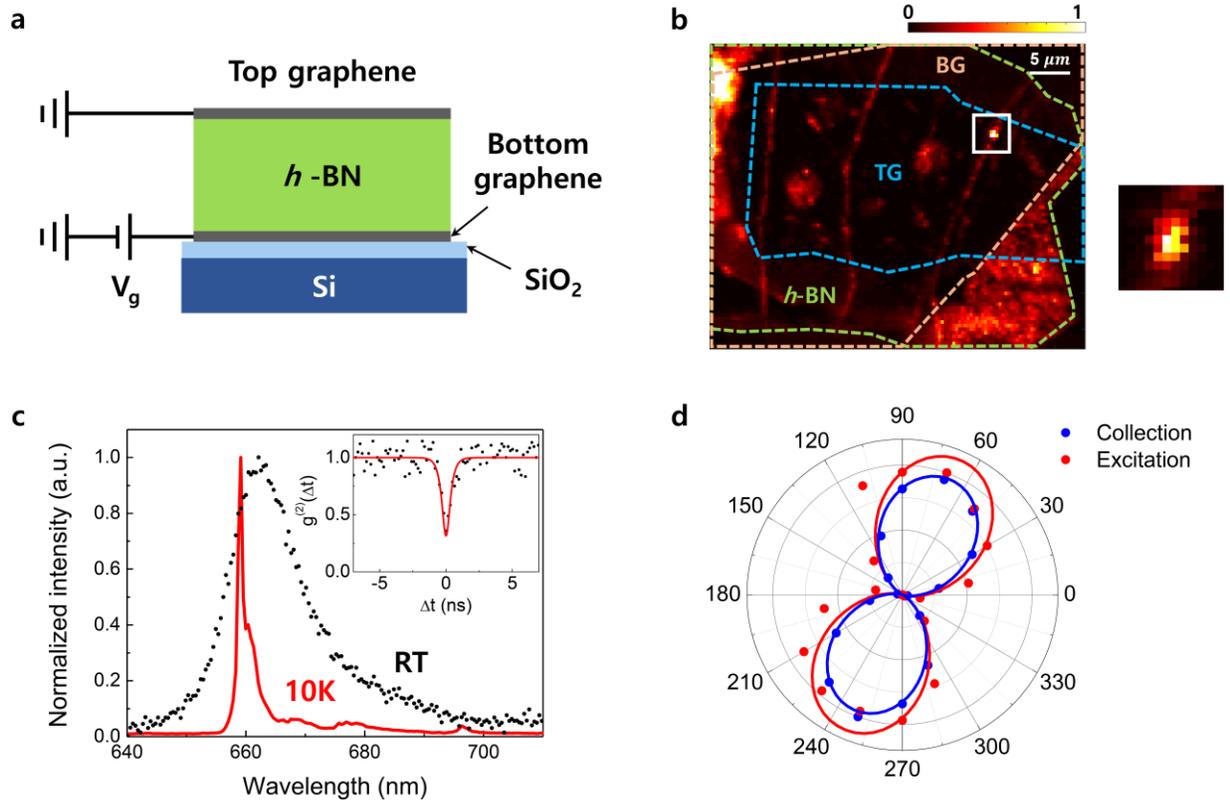

**Fig. 1.** Single photon emission from a point defect in graphene/*h*-BN/graphene heterostructure. **(a)** Device schematics of multi-layer *h*-BN sandwiched by top and bottom few-layer graphene. **(b)** Scanning photoluminescence image of device A measured at 10 K. The squared bright spot shows a localized defect emission. The boundaries of top graphene (TG), *h*-BN and bottom graphene (BG) are shown by dashed lines. **(c)** Photoluminescence spectrum of the emitter shown in **(b)** measured at 10 K and room temperature. In the inset is the second-order correlation ($g^{(2)}(\Delta t)$) function of the emitter showing single photon purity of $g^{(2)}(0) = 0.3$. **(d)** Emission intensity dependence on excitation laser polarization (red dots) and collection polarization (blue dots). The red and blue solid lines are fit lines to $\sin^2(\theta)$.

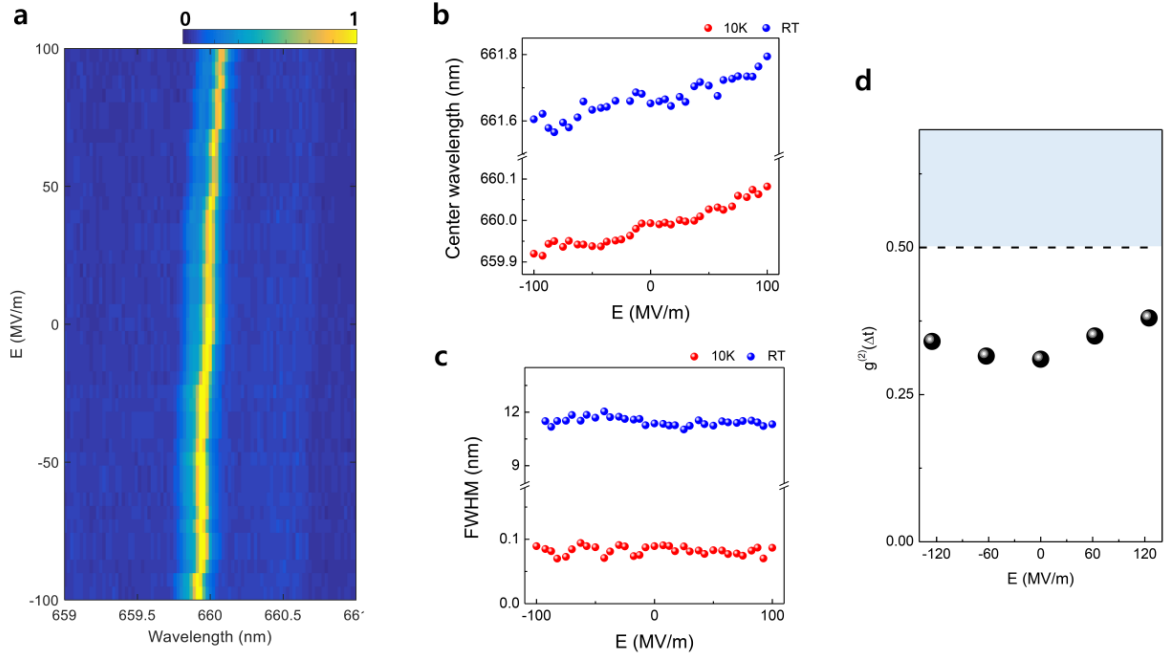

**Fig. 2.** The Stark shift of the single-photon emitter (SPE) shown in Fig. 1. **(a)** Color plot of photoluminescence spectra of the SPE as a function of out-of-plane electric field measured at 10 K. The bright yellow color shows the peak wavelength shifting with the applied electric field. **(b)** The center wavelength shift of the emitter at 10 K (red dots) and room temperature (blue dots) under varying electric fields. **(c)** The FWHM of the emitter at 10 K (red dots) and room temperature (blue dots). **(d)** Second-order correlation function measured at selected fields at 10 K.

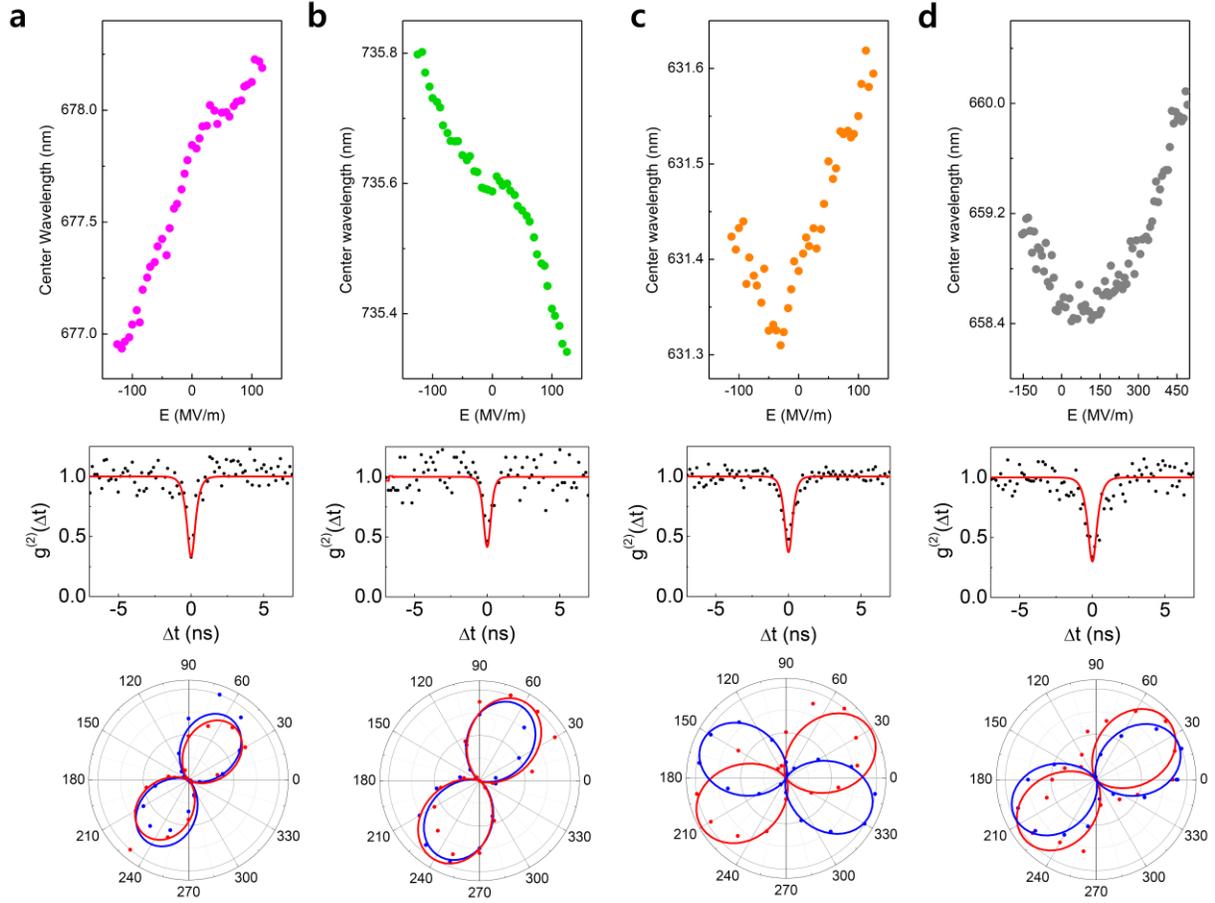

**Fig. 3.** Various types of Stark shifts in different emitters. **(a), (b)** Linear Stark shift observed from two different defects in device A, each showing Δ$\mu$ of 0.9 D (a) and -0.19 D (b), respectively. **(c)** V-shaped Stark shift measured from a defect in device A. The extracted |Δ$\mu$| is 0.24 D. **(d)** Quadratic Stark shift measured from a defect in device B. The extracted Δα and Δ$\mu$ are 149 Å$^3$ and -0.22 D, respectively. The corresponding g$^{(2)}$(Δ$t$) function and polarization dependence are also shown for each emitter.

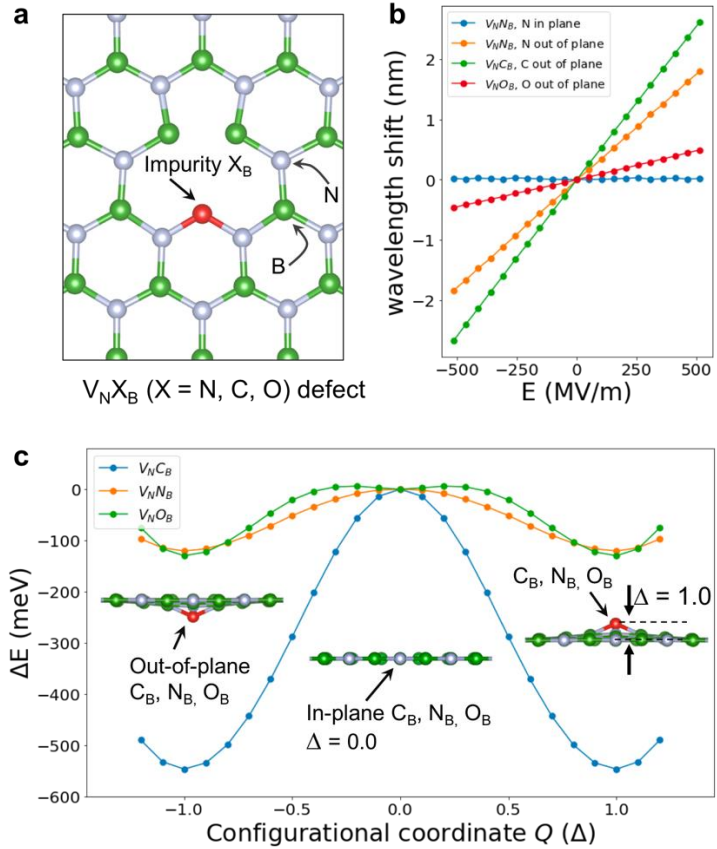

**Fig. 4.** Density functional theory calculations of linear Stark effects. **(a)** Schematic of a $V_NX_B$ defect, which is a pair of an impurity X substituting for boron atom ($X_B$) and its nearest neighboring N vacancy ($V_N$). **(b)** Theoretical ZPL shifts (ZPL($E$) − ZPL($E$=0)) of neutral $V_NX_B$ defects (X = C, N, O) as a function of out-of-plane electric field ($E$). **(c)** Total energy of the $V_NX_B$ (X = C, N, O) defects as a function of configurational coordinate, which is mainly described by out-of-plane displacement of $X_B$. If the impurity atom ($X_B$) resides in-plane ($\Delta = 0$), the symmetry of the defect is $C_{2v}$ that has an inversion symmetry in the out-of-plane direction. However, if the impurity atom moves out-of-plane ($\Delta = \pm 1$), the inversion symmetry is broken, and the defect's symmetry is reduced to $C_{1h}$. The meta-stable $\Delta = 0$ and the ground-state $\Delta = \pm 1$ structures are obtained by full DFT relaxations, while defect structures other than $\Delta = 0$ and $\pm 1$ are generated by linear interpolation and extrapolation.

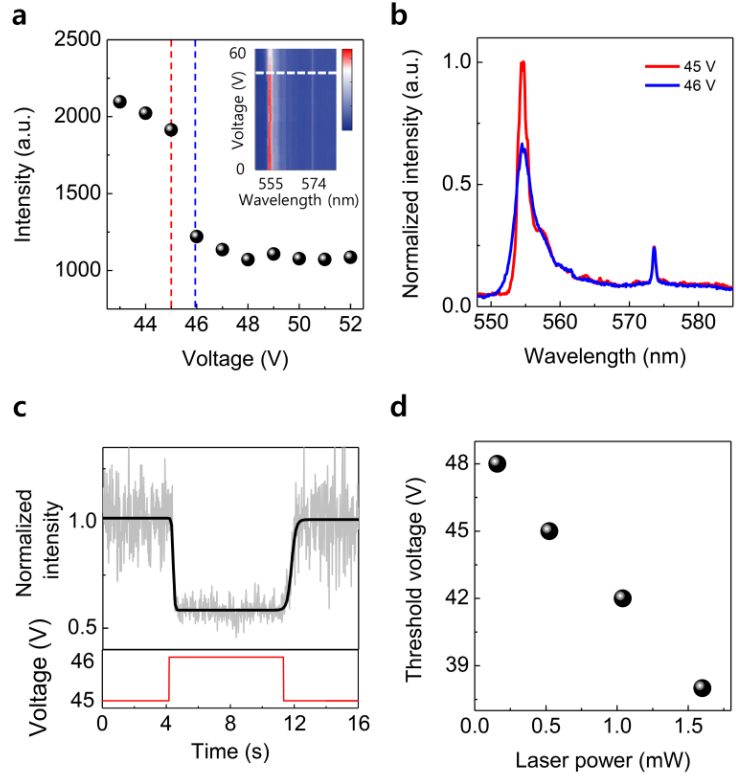

**Fig. 5.** Gate-induced emission modification of a defect in device C. **(a)** Emission intensity of the defect while scanning the gate voltage for the laser power at 0.52 mW. Inset shows the color plot of the spectrum as a function of gate voltage. White dashed line shows the threshold voltage. **(b)** Spectrum of the emitter at two different gates. The peak at 574 nm is *h*-BN Raman shift. **(c)** Time trace of the modified emission intensity (grey, upper) upon the change of the gate voltage (red, lower). The solid black line shows the fitting to a sigmoid function $I = (1 + e^{-k(t-t_0)})^{-1}$. **(d)** The threshold gate voltage measured at different excitation powers.

**ACKNOWLEDGMENT**

This work was supported by the "New Faculty Research Fund" of Ajou University and "Human Resources Program in Energy Technology" of the Korea Institute of Energy Technology Evaluation and Planning (KETEP), granted financial resource from the Ministry of Trade, Industry & Energy, Republic of Korea. (No. 20164030201380). Hosung Seo was supported by the National Research Foundation of Korea (NRF) grant funded by the Korea government (MSIT) (No. 2017R1C1B5077000).

**ASSOCIATED CONTENT**

Supporting Information Available: Device microscope images, raw data of room temperature Stark shift and nonlinear Stark shifts, Stark shift under switched gate and ground electrodes, second-order correlation function at finite fields, details of DFT calculation and defect level diagrams and orbital, and additional data showing the quantum light source characteristics of the defects.

**Table of Contents Graphic**

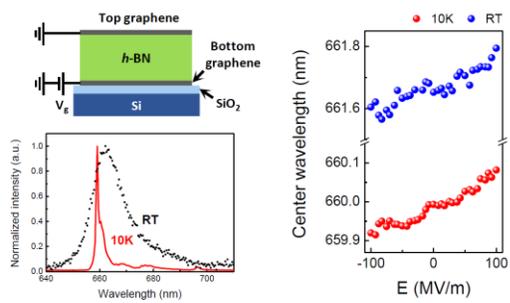